# JNEEG shield for Jetson Nano for real-time EEG signal processing with deep learning

Ildar Rakhmatulin, PhD, PiEEG

**Abstract**.
The article presents an accessible route into the field of neuroscience through the JNEEG device. This device allows converting the Jetson Nano board into a brain-computer interface, making it easy to measure EEG, EMG, and ECG signals with 8 channels. With Jetson Nano is possible use deep learning for real-time signal processing and feature extraction from EEG in real-time without any data transmission. Over the past decade, the proliferation of artificial intelligence has significantly impacted various industries, including neurobiology. The integration of machine learning techniques has opened avenues for practical applications of EEG signals across technology sectors. This surge in interest has led to the widespread popularity of low-cost brain-computer interface devices capable of recording EEG signals using non-invasive electrodes. JNEEG device demonstrates satisfactory noise levels and accuracy for use in applied tasks with machine learning.

**Source** https://pieeg.com/
**GitHub** https://github.com/Pi-EEG/EEG-with-JetsonNano
**YouTube**  https://youtu.be/f3stVQCsfrM

**JNEEG is not medical device**

## 1. Introduction
Electroencephalography stands out as a straightforward and accessible technique for capturing brain signals. It operates by detecting electrical activity in various brain regions through electrodes affixed to the scalp. Widely utilized in both research and clinical settings, EEG signals serve diverse purposes, continually expanding in their applications. Different types of sensors, such as wet and dry electrodes, exist for recording these signals. Wet electrodes, favored for their low impedance typically ranging from 200 kOhm to 5 kOhm after gel application, remain the prevalent choice. Li et al. [1] extensively dissected the merits and drawbacks of wet and dry electrodes, highlighting the nuanced advantages and disadvantages of each type. In our work we used dry electrodes (Ag/AgCl) because opting for dry electrodes eliminates the need for gel application, simplifying EEG measurements.

## 2. Problem statement and motivation
The price range of commercial devices varies significantly, often posing a challenge for researchers due to their high cost. Nowadays, with the surge in interest surrounding neural networks and signal processing techniques, there's a growing curiosity in Brain-Computer Interface devices (BCI) across various application domains. As early as 2016, Meng et al. [3] demonstrated the control of a robotic arm through motor imagery, utilizing neural networks to capture corresponding brain waves via electrodes. This achievement was previously exclusive to invasive methods involving implanted microelectrode matrices for direct brain signal measurement [4]. Invasive techniques involve electrodes making direct contact with the cerebral cortex to capture brain waves locally. Consequently, such methods are cumbersome, expensive, hazardous, and demand highly skilled personnel alongside specialized equipment [5, 6, 7]. Thus, while invasive approaches have shown promise, there's a growing focus on low-cost EEG signal recording devices in this area. Typically, EEG signal readers comprise a microcontroller (processor), a power board, and transmission adapters, the latter often contributing significantly to costs. Hence, this article concentrates on addressing this issue via a special shiel for Jetson Nano is a popular board for using  machine learning for applied task, with our shield we extend the opportunity of the Jetson Nano board to measure biosignals as EEG, EMG EKG and use this information for some applied tasks, as example, control robotics.

## 3. Review of related work

Today, there exists an array of low-cost not medical devices utilized for recording EEG signals. Gunawan et al. [8] utilized a low-cost device to capture EEG signals for various classification tasks, while Ashby et al. [9] employed the same device to classify mental visual tasks. Seneviratna et al. [10] introduced a device transmitting data to a computer via Bluetooth from seven EEG channels and one audio channel. The wireless EEG device outlined by Nithin et al. [11] shares similar limitations. Chapman et al. [12] presented a novel BCI, yet technical details such as component specifications or circuit board diagrams were absent from the paper. Numerous studies have explored employing RaspberryPI for EEG signal reading [20]. For instance, Dillon [13] utilized an ADC (mcp3008) to measure EEG signals for brain injury diagnosis. Ukveris et al. [14], in a study closely related to this article, developed a screen for EEG measurement using Atmega2560 with Arduino. But none of these devices made it possible to use machine learning natively on the device and perform function extraction in real-time. This article addresses introducing a new JNEEG device, a comprehensive self-contained BCI based on Jetson Nano.

## 4. Technical realizations

The leading role in EEG task measurements among analog-digital converters (ADCs) can be placed on the ADS1299 block produced by Texas Instruments. With over a decade in the market, it has consistently been regarded as one of the top units available. What sets this ADC apart from its competitors is its built-in multiplexer. Usman et al. [15] and Rakhmatulin et al. [16] extensively reviewed the capabilities of this ADC and the features of the ADS1299 multiplexer, a discussion that is beyond the scope of this article. Prior to testing, impedance was verified using the ADS1299 function following the methodology outlined in the article [17]. We positioned 8 Ag/AgCl dry electrodes according to the International Electrode Placement System "10-20", selecting the following locations: F7, Fz, F8, C3, C4, T5, Pz, and T6. The general view of JNEEG is presented in Fig.1.

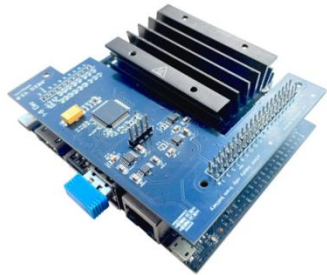

Fig.1. General view of the PCB boards with soldered elements

During the test, the device remained disconnected from the network. We employed a power bank for the purpose of security and to prevent any potential network disruptions. The device was exclusively tested using 5V batteries (and must be tested and used only from 5V batteries). A general view of the in the assembly device configuration and the arrangement of the electrodes is shown in Fig. 2.

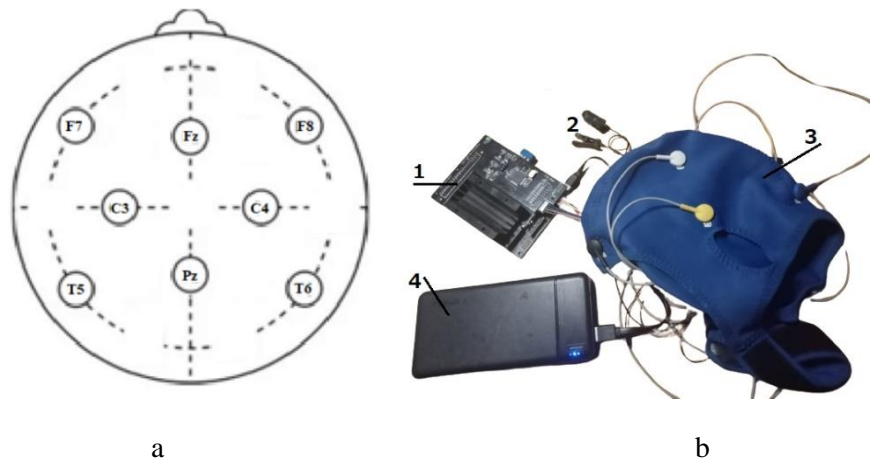

a                                              b

Fig.2. a - Electrode arrangement according to the international 10/20 system , b - General view of the assembled device: 1 –JNEEG, 2 – Clips Electrodes, 3 – EEG cap with, 4 – power battery

## 4. Test device
### 4.1 Artefacts

Internal noise for JNEEG lies within 1 µV. We realized several tests for reading EEG signals for the presence of the simplest artifacts - chewing and blinking. These artifacts are easy to use for testing a developed EEG device [18]. Sheoran [19] detailed how to detect it and how to deal with this artifact. In our experiments, the blinking and chewing artifact was strongly distinguished against the background of the EEG signal (Fig. 3).

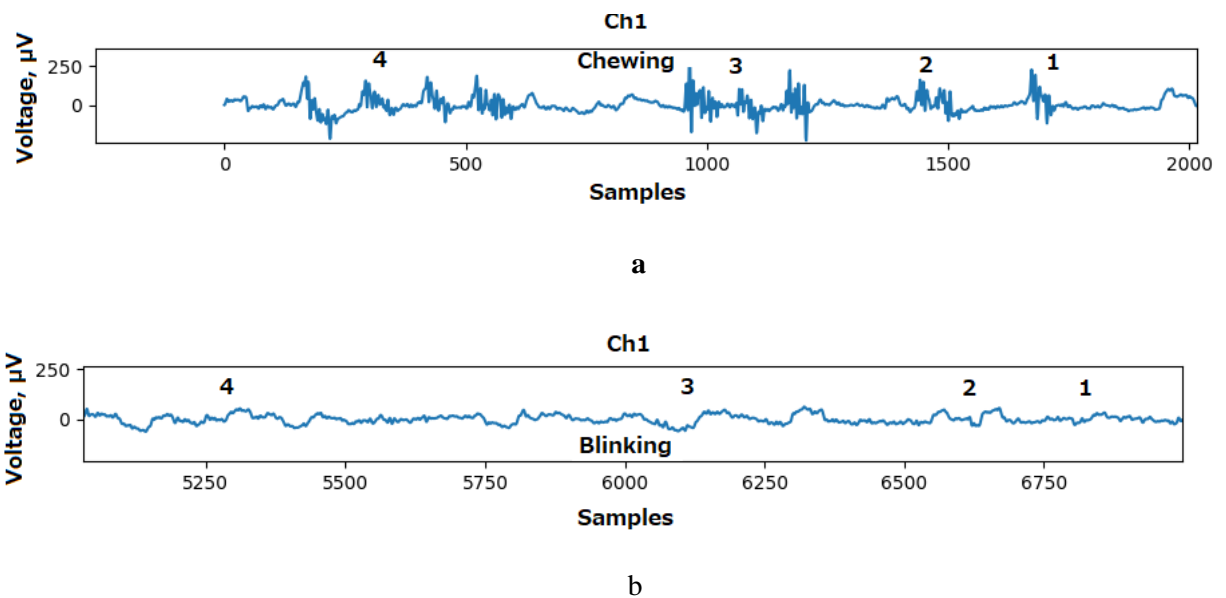

a

b

Fig.3. Artifact test. The process of measuring chewing and blinking artifacts using dry electrodes (Fz). a - Chewing occurred in the following sequence: 4 times, 3 times, 2, and 1 time, and b - the same for the blinking process. The y-axis is the processed EEG signal after passing filter bands of 1-40 Hz in microvolts and with 250 samples per second

### 4.2 Alpha

An alpha wave (α-wave) brain signal, ranging from 8 to 12 Hz with an amplitude typically between 35 and 65 µV, serves as a suitable test for assessing the EEG recording system. Typically, individuals in an awake state with closed eyes, during relaxation, can detect alpha waves. Each EEG signal was measured for a duration of 8 seconds with both eyes closed and open. As anticipated, there was an observed increase in EEG signal amplitude within the frequency band of 8 to 12 Hz. Likewise, a decrease in alpha activity was noted when the eyes were open. These findings align with the anticipated alpha wave pattern in the occipital lobe of the brain, validating the proper design and functioning of the device, Fig.4.

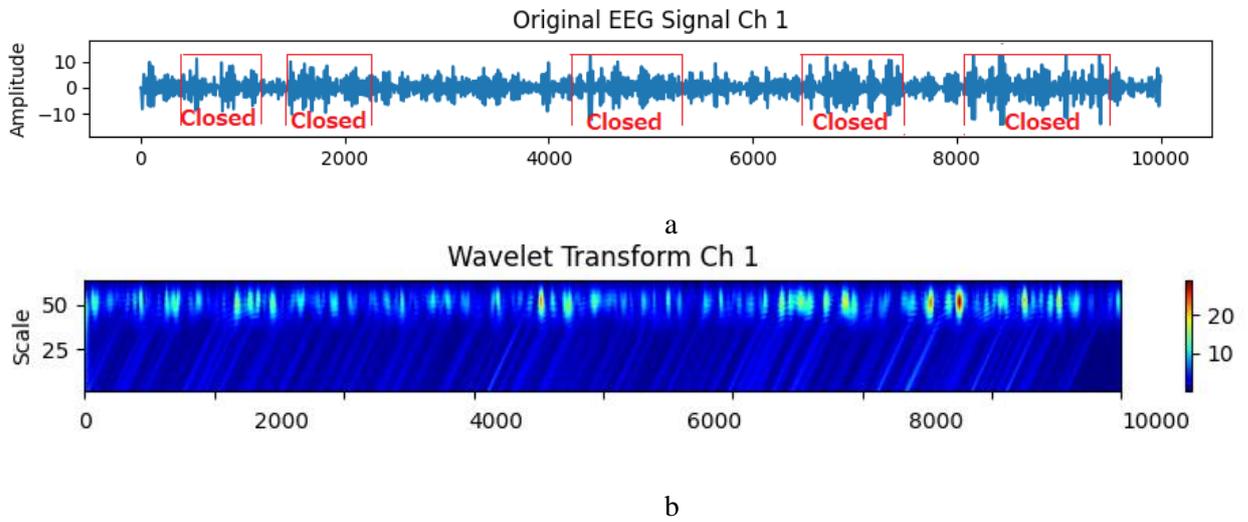

Fig.4. Alpha test. The process of recording an EEG signal from an electrode (Fz) with eyes open and closed, a – voltage visualization, b – wavelet transform. The y-axis is the processed EEG signal after passing filter bands of 8-12Hz in microvolts and with 250 samples per second

## 5. Deep learning opportunity with Jetson Nano

Over the past years, dozens of articles have been published in which researchers used the Jetson nano as a device for computing and feature extracting from EEG signals. In the paper [21], the Jetson Nano was employed for the purposes of data mining and predicting liver disease. The authors assessed the accuracy of the EPILEPSIAE database, one of the most extensive publicly accessible epilepsy datasets, for seizure detection. The proposed system attained a sensitivity of 81.25%. Similarly, in the study [23], epilepsy has already been effectively addressed in practical applications through the utilization of Brain-Computer Interface (BCI) technology in conjunction with Jetson Nano. In paper [24], attention is gauged through a multimodal approach, employing the Jetson Nano neural network alongside Brain-Computer Interface (BCI) technology to assess students' levels of attentiveness. In the paper [22], the authors successfully controlled a wheelchair in real-time with the aid of Brain-Computer Interface (BCI) and Electroencephalography (EEG) technology. In the paper [26], the authors introduced a Convolutional Neural Network (CNN) architecture designed for the classification of reach and grasp actions, focusing on neural correlates from a peripheral perspective. In study [25], also dedicated to data processing, the authors successfully implemented multi-channel continuous wavelet transform on Nvidia Jetson single-board computers to analyze EEG data in real-time.

The Jetson Nano, functioning as a hardware platform, excels in data processing, particularly through machine learning applications and functionalities. Studies demonstrate its efficacy in medical disease detection and external object control applications. However, the external Brain-Computer Interface (BCI) in these endeavors poses challenges for practical implementation, as it introduces complexity in setup and escalates costs. In our context, the connection between Jetson Nano's computational capabilities and its potential as a BCI for biodata collection in one setup offers a promising avenue. This approach can enables Jetson Nano to serve both as a robust computing tool and a biodata acquisition device, facilitating streamlined and cost-effective solutions.

## 5. Conclusion and Discussion

This article highlights the distinctiveness of the presented device by emphasizing its combination of low cost and high accuracy. Additionally, it underscores the uniqueness of the work by confirming measurement accuracy through established methods involving a well-recognized artifact and alpha waves. Furthermore, the article mentions that the noise characteristics of the proposed device align well with those of the ADS1299 from Texas Instruments, validating the quality of the affordable board. It also suggests a promising avenue for future research: practical application as a brain-computer interface. The article asserts that the proposed device ensures high-quality data transfer between the ADC and the processor without introducing any delays, making it suitable for real-time use.

Given the widespread popularity of Jetson Nano, we anticipate that this article will capture the interest of numerous researchers. We express hope that the project will facilitate the accumulation of a substantial dataset, and to that end, we have made the software project in open-source format. We extend an invitation to others to collaborate on data processing efforts.

**CONFLICTS OF INTEREST:** NONE
**FUNDING**: NONE